\documentclass[
reprint,
 aps,
 superscriptaddress,
bibnotes,
]{revtex4-2}

\usepackage{physics}
\usepackage{bbold}
\usepackage{dsfont}
\usepackage{color,newfloat}
\usepackage{graphicx}
\usepackage{dcolumn}
\usepackage{bm}
\usepackage{amsmath,amssymb,amsthm}
\usepackage{mathtools}
\usepackage{multirow}
\usepackage{hhline}
\usepackage[dvipsnames]{xcolor}
\usepackage{hyperref}
\hypersetup{
    colorlinks=true,
    linkcolor=Maroon,
    citecolor=Maroon,
    urlcolor=Maroon
}
\usepackage[caption = false]{subfig}
\usepackage{lineno}

\usepackage{makecell}

\begin{document}


\title{
Deterministic generation of concatenated graph codes from quantum emitters
}

\author{Love A. Pettersson}
\email{love.pettersson@nbi.ku.dk}
\affiliation{Center for Hybrid Quantum Networks (Hy-Q), Niels Bohr Institute, University of Copenhagen, Blegdamsvej 17, 2100 Copenhagen, Denmark.}

\author{Anders S. Sørensen}
\email{anders.sorensen@nbi.ku.dk}
\affiliation{Center for Hybrid Quantum Networks (Hy-Q), Niels Bohr Institute, University of Copenhagen, Blegdamsvej 17, 2100 Copenhagen, Denmark.}

\author{Stefano Paesani}
\email{stefano.paesani@nbi.ku.dk}
\affiliation{Center for Hybrid Quantum Networks (Hy-Q), Niels Bohr Institute, University of Copenhagen, Blegdamsvej 17, 2100 Copenhagen, Denmark.}
\affiliation{NNF Quantum Computing Programme, Niels Bohr Institute, University of Copenhagen, Blegdamsvej 17, 2100 Copenhagen, Denmark.}

\date{}


\begin{abstract}
Photon loss is the dominant noise mechanism in photonic quantum technologies. Designing fault-tolerant schemes with high tolerance to loss is thus a central challenge in scaling photonic quantum information processors. 
Concatenation of a fault-tolerant construction with a code able to efficiently correct loss is a promising approach to achieve this, but practical ways to implement code concatenation with photons have been lacking. 
We propose schemes for generating concatenated graph codes using multi-photon emission from two quantum emitters or a single quantum emitter coupled to a memory; capabilities available in several photonic platforms.
We show that these schemes enable fault-tolerant fusion-based quantum computation in practical regimes with high photon loss and standard fusion gates without the need for auxiliary photons.

\end{abstract}

\date{\today}

\maketitle


\noindent\textbf{Introduction.}
Scalable quantum computing architectures rely on fault-tolerance to mitigate the detrimental impact of physical imperfections encountered during quantum computations~\cite{ShorQEC, KnillQEC}. 
To reach fault-tolerant regimes with near-term and future quantum hardware, it is critical to design fault-tolerant architectures tailored to the native error mechanism and operations for a chosen hardware platform~\cite{IonTrapArch, CVArchOne, CVArchTwo}.
Photon loss and the probabilistic nature of photon-photon entangling gates with linear optics are the dominant noise mechanisms in photonic devices but are typically uncommon for other quantum computing platforms. 
This difference limits the suitability of conventional quantum computing architectures to photonic hardware.
Fusion-based quantum computing (FBQC), a variant of measurement-based schemes~\cite{Oneway, BriegelMBQC, RaussendorfMBQC}, has been developed as a useful framework for developing architectures tailored to photonic hardware~\cite{PsiFBQC}. 
These approaches are based on probabilistic two-qubit parity measurements, so-called \textit{Bell state} or \textit{fusion measurements}~\cite{browne2005} -- operations that can be readily implemented using simple linear-optical circuits acting on single photons~\cite{SegoviaPerc}.
Fault-tolerant computations can be achieved by consuming photonic qubits that are initially part of limited-size entangled states, denoted as \textit{resource states}.
The outcome of such parity measurements is used to process quantum information and, at the
same time, extract syndrome data for error correction~\cite{PsiFBQC}.  
Both photon loss and probabilistic fusion failure effectively act as erasure of parity check outcomes from fusion measurements, which can be directly detected and corrected with standard quantum error correction decoders~\cite{Pymatching, DelfosseUFD}.
Despite significant progress in the design of FBQC architectures with increased tolerance to erasure~\cite{StefanoBen, Passivebias, Psiadaptive}, current fault-tolerant FBQC constructions have photon loss thresholds at rates below $\sim 1\%$, which is extremely challenging to reach in practice.
Concatenation of the resource states with a small inner quantum code \cite{Codeconcat, Tom}, enabling \textit{logical fusions} to fuse logical qubits in the resource states, have been shown to be capable of boosting the resulting photon loss thresholds to $>10\%$, significantly more amenable to current hardware capabilities.
However, schemes for practically generating the concatenated resource states of photons have been lacking, limiting the prospects of photonic quantum computing hardware to loss requirements largely beyond what is achievable with near-term technologies.
Here, we propose protocols for the deterministic generation of concatenated photonic resource states with quantum emitters - single-photon sources with a spin-photon interface~\cite{Uppu2021, LidnerRudolph, CiracSpinPhotonInterface}. 
We show that resource-efficient and practical schemes with only two quantum emitters, or a single emitter coupled to a memory, and sparse spin-spin interactions are sufficient to generate concatenated resource states suitable for fault-tolerant FBQC. 
We optimize over small-scale concatenated resource states that can be generated with such schemes and show that loss thresholds above $4\%$ become achievable already with existing FBQC fault-tolerant models. 
Our schemes are flexible and amenable to a number of different types of quantum emitters, including quantum dots~\cite{Schwartz2016, Istrati2020, Appel2022, meng2023deterministic, Coste2023}, colour centers~\cite{vasconcelos2020}, atoms~\cite{Yang2022, Thomas2022, thomas2024fusion} and ions~\cite{NetworkTrappedIonsMemory, MonroeIonSpinPhotonInterface}, evidencing their relevance for developing quantum photonic technologies in a variety of platforms and for a broader range of applications, e.g. quantum networking~\cite{QuantumRepeatersBrigel, QuantumRepeatersReview}.

\begin{figure*}[t]
    \centering    \includegraphics[width=1\textwidth]{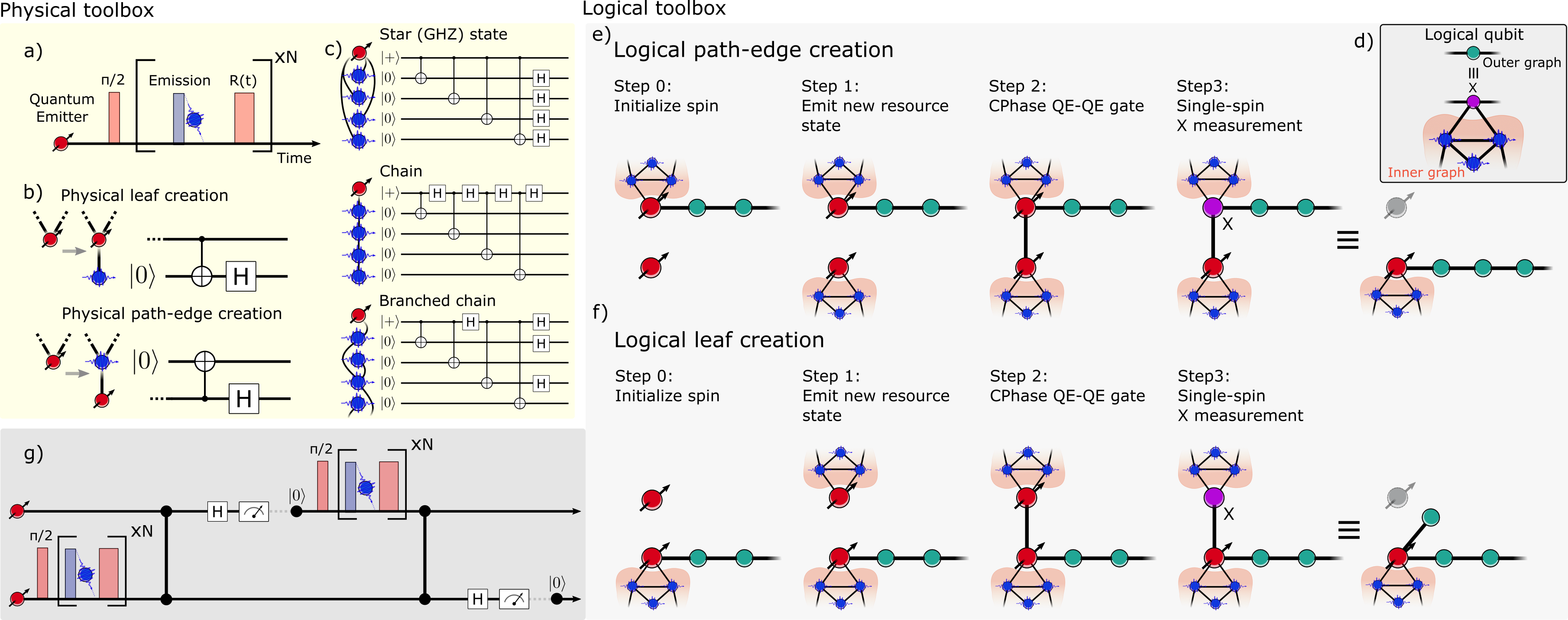}
    \caption{
   \textbf{Generation schemes for physical and logical resource states with quantum emitters.}
    a) Physical resource states of entangled photons can be generated via pulse sequences driving a spin-photon interface. 
    The sequences are composed of alternating pulses driving the spin (shown in red) and pulses exciting the quantum emitter for photon emission (in blue).
    b) The generation of each photon has a graph description and is equivalent to either the creation of a leaf (top) or a path-edge (bottom) depending on the Clifford operations performed on the spin and photon.   
    c) Sequences of single-qubit gates and photon generation create resource states locally equivalent to three graph classes: star graphs (i.e. GHZ states), chains, and branched chains.  
    d) Graph states of logical qubits (shown in green) encoded in an inner code correspond to having an inner graph appended to each node of the outer graph, with the outer nodes representing virtual qubits considered to be measured in $X$ with an outcome $+1$.
    Steps for the implementation of logical graph state generation are shown in e) for the path-edge creation and f) for the leaf creation. 
    Note that, in f) the spin generating the graph code and the spin being re-initialized (measured in X) is the same, while in e) one spin generates the graph code and the other is re-initialized.
    These steps perform, using two quantum emitters, the same operations as in b) but at the logical level.
    g) Pulse sequences and spin-spin gates associated with the logical graph operations.
    In all figures, spin qubits are shown in red, physical photonic qubits are shown in blue, logical qubits are shown in green, and virtual qubits in purple.
    }
    \label{fig:1}
\end{figure*}

\ \\
\noindent\textbf{Graph state generation with quantum emitters.}
Quantum emitters are a platform with strong potential to implement photonic architectures as they enable the deterministic generation of entangled resource states of photons~\cite{LidnerRudolph}. 
These systems, which include quantum dots~\cite{Schwartz2016, Istrati2020, Appel2022, meng2023deterministic, Coste2023}, atomic systems~\cite{Yang2022, Thomas2022, thomas2024fusion, meng2023fusion}, and color centers~\cite{vasconcelos2020}, can do so via pulse sequences driving a spin-photon interface interleaved with photon emission~\cite{LidnerRudolph}, as depicted in Fig.~\ref{fig:1}a.
The states generated through this process can be conveniently described as graph states, where each vertex of the associated graph representation is a qubit initialized in the $\ket{+}=(\ket{0} + \ket{1})/\sqrt{2}$ state \footnote{Here $\ket{+}$ is the plus eigenstate of the Pauli $X$ matrix represented in the Pauli $Z$-eigenstates $\ket{0}$ and $\ket{1}$ which can be represented by vectors $\big(\begin{smallmatrix}1 \\0\end{smallmatrix}\big)$ and $\big(\begin{smallmatrix}0 \\1\end{smallmatrix}\big)$, respectively.} and edges are controlled-$Z$ entangling operations~\cite{hein}. 
In this description, it is useful to distinguish between vertices of the graph associated with spin qubits (pictured in red the figures) and photonic qubits (in blue), as they play different roles in the resource state generation. 
For a single emitter, the entanglement structure generated through general pulse sequences~\footnote{Strictly speaking, in the graph-based description of resource state generation considered here, we are restricting ourselves to the case where all gates applied to the spin and photons are Clifford gates. As far as we are aware, it is not currently known whether non-Clifford gates on the spin could enable the generation of graph resource states outside the set generated by the graph operations considered here.} as in Fig.~\ref{fig:1}a can be represented, up to local operations on the photons, via simple graph operations. 
The generation of each photon can be described as either (1) a \textit{leaf creation}, i.e. the generation of a photonic vertex of degree 1 connected to the spin qubit emitting the photon, or (2) a \textit{path-edge creation}, which we define as the creation of a leaf attached to the spin qubit followed by swapping the two vertices so that the leaf is now the spin attached to a photonic qubit \cite{StefanoBen}.
These operations are depicted in Fig.~\ref{fig:1}b together with the associated quantum circuit representation.
Physically, the choice between (1) and (2) for a given photon emission is determined by the single-qubit gate performed on the spin (the $R$ rotation in Fig.~\ref{fig:1}a) before the photon emission: if the identity is performed then the photon generation is described by (1), if a Hadamard $H$ gate is performed then (2) is obtained.
Performing other Clifford gates leads to operations equivalent to these after suitable basis transformations.
The leaf and path-edge creation operations are the building blocks for generating resource states with a single quantum emitter. 
Sequences of them can generate the following classes of graphs, represented in Fig.~\ref{fig:1}c: star-shaped graphs (locally equivalent to GHZ states), chain graphs shown, and branched-chain graphs (also known as ``caterpillar graphs''). 
All other resource states that can be generated with a single quantum emitter are equivalent to these graph structures up to local operations~\cite{StefanoBen}.

\ \\
\noindent\textbf{Schemes for code concatenation.}
We now illustrate how, by using two coupled quantum emitters, it is possible to generate concatenated resource states of photons deterministically.
Code concatenation describes having each qubit forming the codewords of a code being encoded in another code — a standard approach in quantum error correction~\cite{Codeconcat}.
In other words, each qubit in the original code, which we call the \textit{outer code}, is itself a logical qubit (shown in green in all figures) encoded in an inner code. 
In terms of graphs, concatenation can be seen as embedding each node of the original graph, which we call the \textit{outer graph}, with an \textit{inner graph} that represents the concatenated code, and considering the vertices of the outer graph as virtual qubits (purple) measured in the Pauli $X$ basis and with +1 outcome obtained (as depicted in Fig.~\ref{fig:1}d)~\cite{hein, PsiFBQC, Tom, Codeconcat}.
Virtual qubits do not have to exist in practice but are useful in describing the concatenated graph; only the qubits in the inner graphs needs to be physical. 
However, in our approach, we will consider them to spin qubits, which we assume can be measured with high measurement probability.
To describe the generation of concatenated resource states we show how two coupled quantum emitters can perform the same graph building operations as the single emitter case described above, i.e. leaf and path-edge creation, but with the generated graph vertices now representing logical qubits.
The schemes for the creation of a \textit{logical leaf} and a \textit{logical path-edge} are described in Fig.~\ref{fig:1}e and Fig.~\ref{fig:1}f, respectively.
The protocol starts by generating the inner code for a first logical graph vertex with a quantum emitter and then adding more logical vertices to the outer code by recursively applying steps 0-3 in Fig.~\ref{fig:1}e-f) for logical leaf and path-edge creation operations.
Such operations are ``logical'' equivalents of the physical building blocks, which can now be used to generate the same entanglement structures described for a single-quantum emitter but concatenated with an inner graph code. 
With these schemes, the inner code and the outer code are thus in the same class of graph states; those locally equivalent to the graphs in Fig.~\ref{fig:1}c.
The schemes can be understood as sequential generation of resource states from one emitter, alternating between the two, followed by a single controlled-$Z$ operation between the emitters and measurement of one emitter in Pauli $X$ to perform concatenation.
The choice between logical leaf or logical path-edge creation is simply determined by which of the quantum emitters is measured in $X$ after the spin-spin gate.
We remark that the single-qubit spin $X$ measurement should provide $+1$ outcome.
However, if $-1$ is obtained then it can be taken into account by a sign flip on the logical $\overline{X}$ operator of the inner code which can be tracked classically, equivalently to Pauli frame updates in topological codes~\cite{Tom, SurfaceCodeReview}.

\begin{figure}[]
\includegraphics[width=0.49\textwidth]{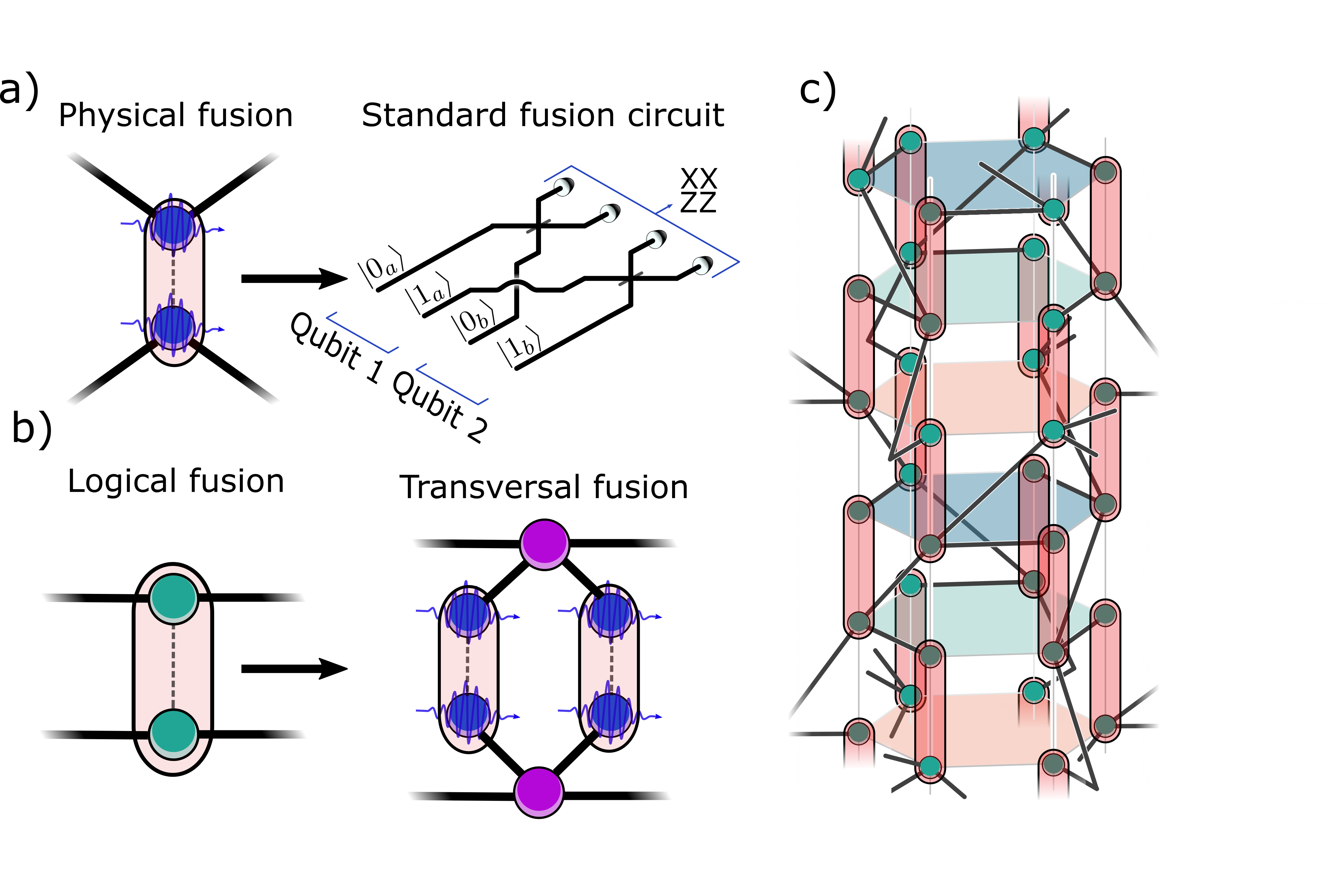}
\caption{\textbf{Fusion gates and the FFCC lattice}. a) Illustrates a physical fusion measurement between two photonic qubits and its optical circuit realization. b) A logical fusion between two graph codes, which is performed by transversely fusing identical code qubits. 
c) Construction of the FFCC lattice for fault-tolerant fusion-based quantum computation from Ref.~\cite{StefanoBen} using the concatenated logical branched chains generatable with our scheme.
\label{fig:results}}
\end{figure}

An analog construction of the above scheme can also be used for the case where a single quantum emitter is coupled to a long-lived quantum memory, as is the case for example in nuclei-electron spin systems in diamond color centers~\cite{Fuchs2011} or networked trapped ion architectures~\cite{NetworkTrappedIonsMemory, MonroeIonSpinPhotonInterface}.
In such cases, the protocol starts by generating a graph code with the emitter and then appending it to the memory with a SWAP two-qubit gate gate
Furthermore, the schemes are modified by implementing a SWAP gate after the controlled-$Z$ operation between the quantum memory and the quantum emitter for every the path-edge creation.
With the SWAP gate, the measurement and reinitialization are always performed on the short-lived quantum emitter spin while the long-lived quantum memory stores coherence.
Note that, because the concatenated state is a graph state itself, other schemes proposed for generating arbitrary graph states with multiple coupled quantum emitters could also be considered for their generation~\cite{Li2022}.
We show in Appendix.~\ref{app:appendixname1} that the protocol proposed here, targeting specifically the generation of concatenated codes, is significantly more efficient than previous general approaches in terms of hardware requirements.

\begin{figure*}[]
\includegraphics[width=1\textwidth]{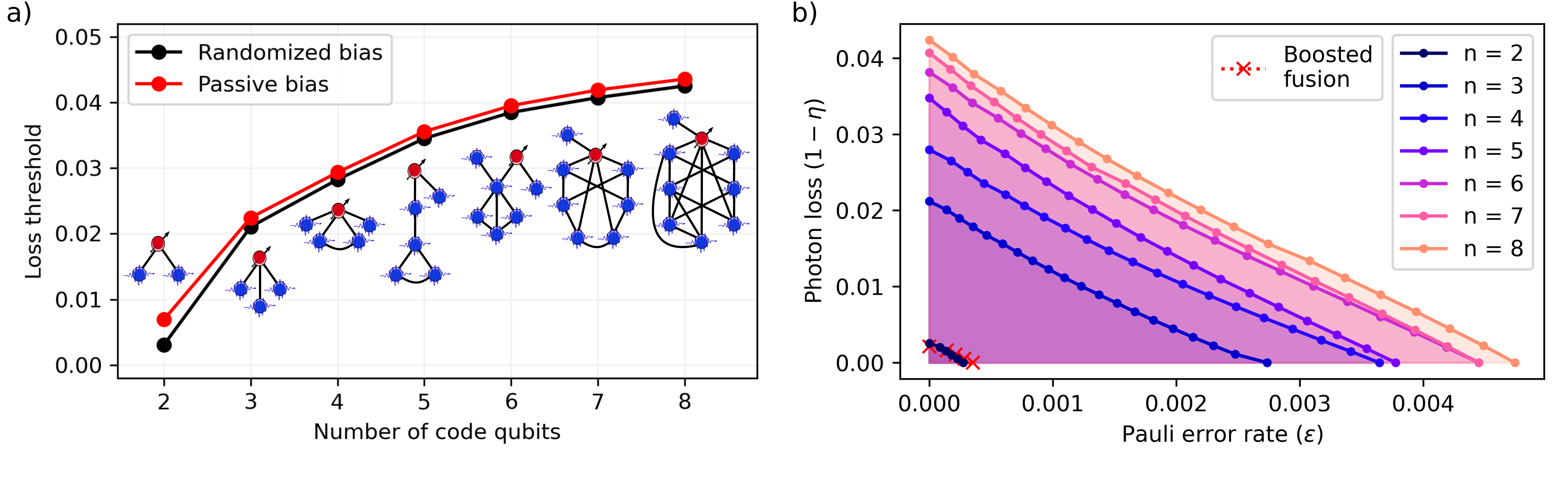}
\caption{\textbf{Photon loss and Pauli error thresholds}. a) The photon loss threshold for the optimized graph codes ranging from two to eight code qubits, for both passive and randomized bias configurations. 
The graph states corresponding to the optimized inner codes (for $\overline{ZZ}$, see Appendix.~\ref{app:graph_lib} for the duals) are shown as inset. b) Fault-tolerant regions for photon loss ($1- \eta$) and Pauli errors ($\epsilon$) obtained using the optimized inner codes in a) under a randomized bias configuration and using phenomenological noise models. Shaded regions indicate the correctable region and all the physical fusions between code qubits are standard fusions with $p_{\text{fail}} = 1/2$ failure probability. The correctable region for boosted fusion ($p_{\text{fail}} = 1/4$) and no code concatenation is also shown for comparison (cross markers) and is very close to the results for $n=2$.  \label{fig:results_error_loss}}
\end{figure*}

\ \\

\ \\ 
\noindent\textbf{Fusion-based quantum computation with logical resource states.}
A prominent application of deterministically generated photonic resource states is FBQC~\cite{PsiFBQC}, a variant of the measurement-based approach to fault-tolerant quantum computing~\cite{Oneway, BriegelMBQC, RaussendorfMBQC} that is particularly suited for photonic hardware. 
In the FBQC model, a network of entangled resource states of photonic qubits is consumed through pairwise entangling measurements, so-called \textit{fusion gates}, to perform quantum computation fault-tolerantly.  
As depicted in Fig.~\ref{fig:results}a, fusion gates are native linear-optical two-qubit operations that correspond to probabilistic entangling measurements, implementable with simple and low-depth optical circuits~\cite{browne2005}.
The fusion of two photonic qubits consumes them but provides joint parity measurement outcomes which can be used to construct syndrome graphs for, e.g., topological fault-tolerant codes~\cite{PsiFBQC, StefanoBen}.
For example, standard fusion gates such as the one depicted in Fig.~\ref{fig:results} implement probabilistic Bell measurements consuming two qubits to provide the outcomes of their joint Pauli operators $XX$ and $ZZ$~\cite{browne2005}.
Upon failure of the fusion gate, which for standard fusion gates happens with probability $p_\text{fail} = 50\%$, the outcome for one of the two parity operators is erased.
Moreover, if any of the photons involved in the fusion are lost, both $XX$ and $ZZ$ outcomes are erased.
Because the fusion outcomes form syndromes in FBQC fault-tolerant architectures, fusion failure and qubit loss, dominant noise sources in photonic quantum hardware, can both be simply described as erasures in the syndrome graph, equivalent to single-qubit erasures in standard measurement-based architectures~\cite{RaussendorfWithLoss}.
Since a fusion failure basis has to be picked, that is, which parity is recovered when the fusion fails but no photon is lost, the erasure rate for $ZZ$ and $XX$ are biased, i.e. $p_{\text{erase}}(ZZ) \neq p_{\text{erase}}(XX)$.
As the $XX$ and $ZZ$ parity outcomes go to different syndrome graphs that are decoded separately, both with
an erasure threshold $\Tilde{p}_{\text{erase}}$ set by the code, fixing the failure basis over the full fusion network the fault-tolerant properties of the code are bounded by the worst erasure rate.

One way of dealing with the bias is uniformly distributing the failure basis between $XX$ and $ZZ$ over the full fusion network, which we call \textit{randomized biased} noise model~\cite{Passivebias,Psiadaptive}.
Assuming the randomized biased model, the erasure rate for both $XX$ and $ZZ$ is $p_{\text{erase}} = 1 - (1-p_{\text{fail}}/2)\eta^{2}$ where $\eta$ is the transmission efficiency for each photon.
With this noise model, the erasure rate that
current FBQC fault-tolerant architectures can tolerate is below the minimum rate obtainable with standard fusions~\cite{PsiFBQC, StefanoBen}. 
A key challenge for FBQC architectures is thus designing fusion schemes to improve the erasure rate to be within the tolerable range. 
One possible approach is to use boosted fusion circuits~\cite{GriceBoost, EwertBoost}, which increase the success probability of fusion gates by using additional ancillary photons.
However, boosted fusions have a very low tolerance to photon loss as the loss of any of the fused or ancillary photons also results in the erasure of both fusion outcomes, resulting in photon loss thresholds $\lesssim 1\%$ for current architectures~\cite{PsiFBQC, StefanoBen}. 
To surpass these limitations, the use of logical resource states has been proposed, where \textit{logical fusions} are performed between qubits encoded via concatenation of the resource state with an inner code~\cite{PsiFBQC, hilaireLossBSM}.
As depicted in Fig.~\ref{fig:results}b, logical fusions measure joint parities of logical operators of the inner codes.
In particular, we will here focus on $\overline{XX}$ and $\overline{ZZ}$ logical fusions between pairs of logical qubits encoded in identical inner graph codes, 
where pairs of physical qubits from each inner code are fused transversely using $XX$ and $ZZ$ fusion measurements, as illustrated in Fig.~\ref{fig:results}b.
The logical erasure rate for the logical parities $\overline{XX}$ and $\overline{ZZ}$ is the probability that they are not recovered 
from the physical fusion outcomes and can be calculated analytically for small inner codes (see Appendix.~\ref{app:erasuredecoder}). 

\ \\
\noindent\textbf{Fault-tolerance performance of FBQC with logical resource states from quantum emitters.}
We analyze the performance improvement in fault-tolerant FBQC architectures that can be achieved with logical resource states generatable with two quantum emitters in the protocols described above. 
In our analysis, we focus on the topological FBQC scheme based on foliated Floquet color codes (FFCC) from Ref.~\cite{StefanoBen}, which is a construction with good fault-tolerance and which requires resource states amenable to quantum emitters. 
In particular, this scheme proceeds by fusing branched chains which, as shown in Fig.~\ref{fig:1}c, are graph states that can be directly generated from a single quantum emitter.
As a consequence, using the protocols described here, logically encoded branched chains can be generated using two emitters (or an emitter and a memory) and employed to implement the same FBQC scheme using logical fusions between the inner codes.
With two emitters, the inner codes that can be used in our scheme are also represented by the class of graph states generatable from a single quantum emitter. 
Here, we analyze the performance in terms of photon loss threshold for the FFCC model with logical fusions for all possible inner codes with up to 8 physical qubits that can be generated with a single quantum emitter \footnote{Here, all graph codes we consider derive from all the graph states which can be generated by a single emitter with the emitter qubit measured in $X$ for code concatenation.}.
Logical fusions are considered to be implemented transversely with standard physical fusions with $50\%$ success probability (providing $XX$, $ZZ$ parities). 
The results optimized overall inner codes, and the associated optimal inner codes, are shown in Fig.~\ref{fig:results_error_loss}a. 

In general, the inner codes can have a bias in the failure rates in the two parities, i.e. $p_{\text{erase}}(\overline{ZZ}) \neq p_{\text{erase}}(\overline{XX})$, depending on the code structure.
In such cases, however, it is always possible to generate a dual inner code that swaps the erasure rates $p_{\text{erase}}(\overline{ZZ}) \leftrightarrow p_{\text{erase}}(\overline{XX})$ of the original code by rotating the spin with a local Clifford on the quantum emitter before measurement. 
This effectively swaps logical operators of the inner code (see Appendix~\ref{App:dual_codes} for more details).
For a given inner code, uniform failure probability in the $\overline{ZZ}$ and $\overline{XX}$ bases of the logical fusion can thus be obtained by randomizing the choice to encode in the code itself or in its dual for every pair of fused logical qubits.
This configuration corresponds to the randomized biased model discussed above~\cite{PsiFBQC}, and the associated optimized loss thresholds are shown in the dark line of Fig.~\ref{fig:results_error_loss}a.
With inner codes with eight physical qubits, we reach a photon loss threshold of approximately $4.4 \%$.
This is a significant increase to the $0.52 \%$ loss threshold achieved in Ref.~\cite{StefanoBen}.

Alternatively, the native bias in the logical fusions can be exploited to tailor weights in the syndrome lattice to restrict errors in 2-dimensional cuts of the 3-dimensional code, which can significantly enhance the noise tolerance~\cite{Passivebias, Psiadaptive, StefanoBen}.
Programming biases can again be implemented by choosing whether to use the inner code or its dual for each logical fusion.
Considering passive biasing patterns (see Appendix~\ref{App:erasure_bias} for more details) for the FFCC code as in Ref.~\cite{StefanoBen} and optimizing over all inner codes available with our scheme for two emitters, we obtain a slight improvement in loss thresholds (light line in Fig.~\ref{fig:results_error_loss}a).
The optimized graph codes for passive and randomized bias coincide, so only one graph code for a given number of code qubits is illustrated in Fig.~\ref{fig:results_error_loss}a).
The redundancy of the logical qubit encoding not only provides loss tolerance but can also simultaneously protect from other qubit errors arising from imperfect gates and measurements on the physical qubits. 
Techniques for decoding and analyzing the performance of general graph codes in the presence of both loss and errors have been developed in Ref.~\cite{Tom}, which we adapt here for analyzing error-corrected logical fusions (see details in Appendix~\ref{app:error_decoder}).
For simplicity, in our analysis, we assume randomized bias in the logical fusions, and as an error model we use the single qubit depolarizing error channel
\begin{equation}
    \mathcal{E} (\epsilon) = (1-\epsilon)\rho + \frac{\epsilon}{3}\sum_{\sigma \in \{ X, Y, Z\}}\sigma \rho \sigma,
\end{equation}
where $\epsilon$ is the single-qubit error rate and $\rho$ is a single qubit density matrix

In Fig.~\ref{fig:results_error_loss}b we show the fault-tolerant regions in the presence of both qubit errors and losses for the loss-optimized codes with up to eight qubits.
For inner codes with eight qubits, we obtain a maximal error rate threshold of approximately $0.47\%$.
For comparison, we show in the same figure also the analog fault tolerant region for a physical construction of the FFCC model where the improvement in fusion success probability is performed via using boosted fusions rather than logical encodings (75\% success probability as in Ref.~\cite{StefanoBen}, reported as \textit{boosted fusion} in Fig.~\ref{fig:results_error_loss}b).
Boosting the fusion two times gives six additional photons to reach a fusion success probability of 87.5\% can increase the tolerance to a loss threshold of $\sim 1 \%$.
Boosting further with additional photons only lowers the threshold due to increased losses with multiple photons.
Thus, the logical encodings perform significantly better than boosted fusion already with small codes of only 3 physical qubits.

\ \\
\noindent\textbf{Discussion.}
We have constructed a deterministic and resource-efficient generation scheme for concatenated photonic resource states, which requires only two emitters and a single round of spin-spin gates per logical qubit. 
This scheme is amenable to a variety of emitter-based platforms as well as memory-emitter systems and performs significantly better in terms of hardware requirements than previous approaches for graph state generation with coupled quantum emitters (see Appendix~\ref{app:appendixname1}).
We have analyzed how such schemes can improve the fault tolerance in FBQC architectures, showing that loss thresholds of $4.4 \%$ can be achieved already with small concatenated codes and current FBQC constructions.
This is significantly larger than the $0.52 \%$ threshold achieved in Ref.~\cite{StefanoBen}, and brings FBQC hardware requirements for quantum emitters much closer to current technological capabilities.
Several techniques can be used to further improve this tolerance, including 1) implementing the logical fusion using adaptive strategies rather than transversely~\cite{Tom}, 2) implementing adaptive biasing to reduce noise transmission in FBQC codes rather than the randomized or passive bias approaches considered here~\cite{Psiadaptive}, 3) constructing FBQC codes with improved tolerance to erasures~\cite{beyondfoliation, Matthias}. 
Furthermore, if more than two interacting emitters are available, the scheme can be readily generalized for the generation of concatenated codes with a larger class of inner codes with improved noise tolerance and more complex logical resource states.


\ \\
\noindent\textit{Acknoweldgements.}
We are grateful to M.C. Löbl, T. Bell, B. Brown, and S. Chen for fruitful discussions. 
We are grateful for financial support from Danmarks Grundforskningsfond (DNRF 139, Hy-Q Center for Hybrid Quantum Networks). L.A.P. acknowledges support from Novo Nordisk Foundation (Challenge project “Solid-Q”). S.P. acknowledges funding from the Marie Skłodowska-Curie Fellowship project QSun (nr. 101063763), from the Villum Fonden research grants No.VIL50326 and No.VIL60743, and support from the NNF Quantum Computing Programme.


\bibliography{biblio.bib}


\newpage 
\clearpage

\pagenumbering{arabic}

\appendix

\renewcommand{\thesection}
{\Alph{section}}

\renewcommand{\thefigure}{A\arabic{figure}}
\setcounter{figure}{0} 

\renewcommand{\thetable}{A\arabic{table}}
\setcounter{table}{0}


\section{Comparsion to existing graph generation schemes}
\label{app:appendixname1}
\begin{figure}[h!]
\includegraphics[width=0.53\textwidth]{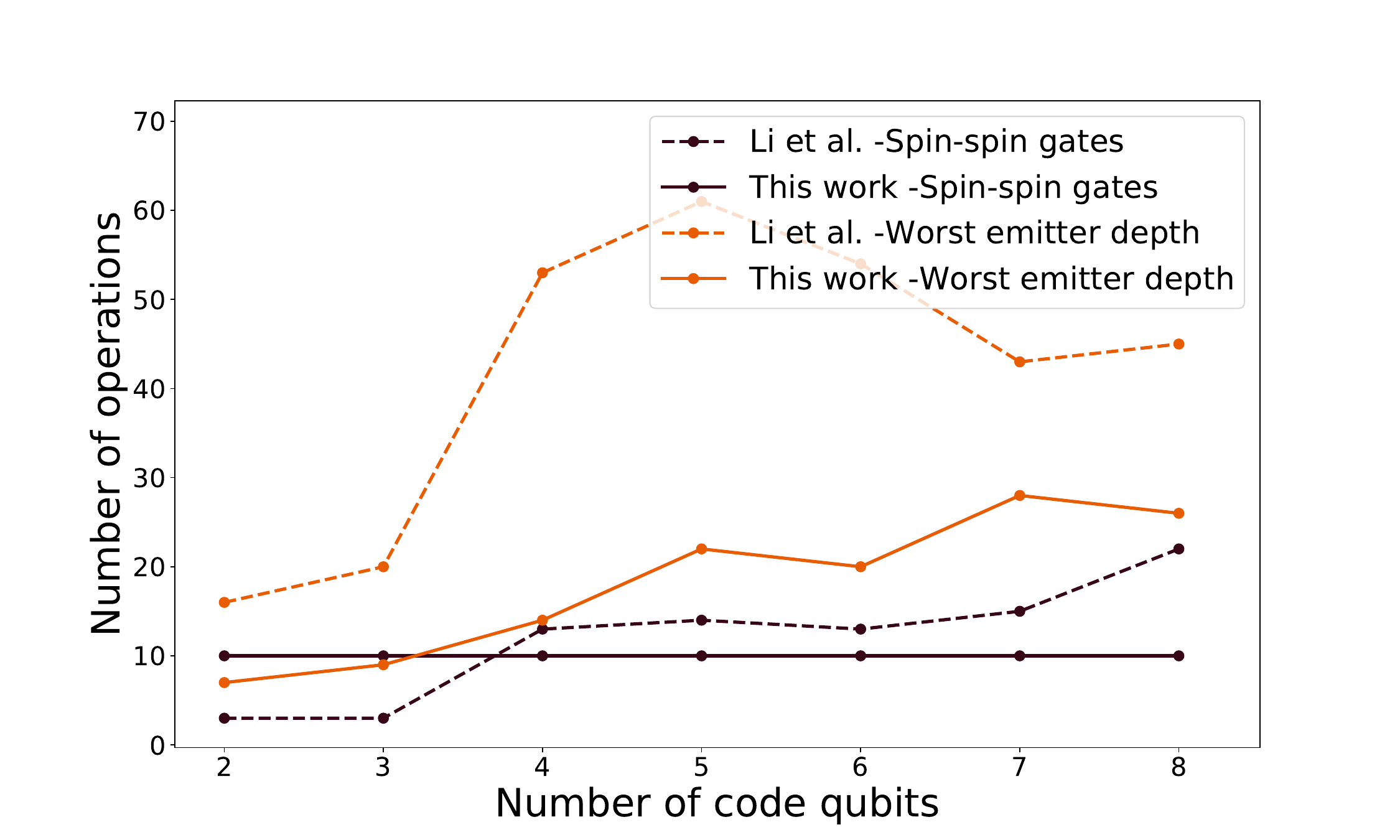}
\caption{\textbf{Requirements for making a ten-qubit concatenated branched linear chain}. A comparison between the generation protocol proposed here and the generation protocol of Li et al. given in Ref.~\cite{Li2022} for a ten-qubit concatenated branched linear chain. The plot shows the scaling of the number of spin-spin gates between emitters and the maximum emitter depth as a function of the number of code qubits for both methods. Note that for the protocol present here, the number of spin-spin gates solely depends on the size of the outer code, which is why it is constant. \label{fig:operation_comparsion}}
\end{figure}

In the main text, we introduce a scheme for generating branched linear resources state concatenated with graph codes that can be generated with a single quantum emitter.
While there already exist protocols to generate arbitrary graph states using a collection of interacting emitters \cite{Li2022}, we show that the protocol presented here is more efficient (i.e. requires fewer operations) than the algorithm in Ref.~\cite{Li2022}. 
Furthermore, the protocol presented here incorporates a convenient reinitialization of the emitters after a fixed number of operations, which is not necessarily integrated in Ref.~\cite{Li2022}.
This can be favorable since the reinitialization acts as a reset of the emitters coherence time.
In Fig.~\ref{fig:operation_comparsion} we compare solutions for a ten-qubit branch linear resource state concatenated with the optimized graph codes in Fig.~\ref{fig:results_error_loss}a in the main text.
When generating solutions with the algorithm in Ref.~\cite{Li2022} we implement the heuristic time ordering of photons introduced in that work, which works well as all solutions, independent of the inner code size, only require two emitters.
In the comparison, we are mainly interested in the number of spin-spin gates between emitters, and the maximum emitter depth.
Here, the emitter depth is defined as the number of operations applied to the emitter before it is measured and reinitialized.
Comparing the two protocols, the largest contribution from the concatenation protocol presented here is the number of gates performed before reinitializing the emitter.
However, we also see a slight decrease in spin-spin gates as the code size grows larger ($n > 3$).

\section{Dealing with erasure bias in FBQC}
\label{App:erasure_bias}
Since a failure basis (i.e. the single qubit basis that the photons are measured in when the fusion fails but no photon is lost) has to be picked for the standard fusion, the erasure rate of the different parities are biased, i.e. $p_{\text{erase}}(ZZ) \neq p_{\text{erase}}(XX)$.
As the $XX$ and $ZZ$ parity outcomes go to different syndrome graphs that are decoded separately, both with
an erasure threshold set by $\Tilde{p}_{\text{erase}}$ as they are self-dual, fixing the failure basis over the full fusion network the worst erasure rate bounds the fault-tolerant properties of the code.
However, there are methods to deal with this bias, and here, we consider two such methods.
In the first method, which we call \textit{randomized bias}, the failure basis is uniformly distributed between the two parities over the whole fusion network.
This is achieved by randomly changing between bases, which sets the erasure rates for both parities to the average of the two $\overline{p}_{\text{erase}} = \frac{p_{\text{erase}}(XX) + p_{\text{erase}}(ZZ)}2$, which for standard fusion is $\overline{p}_{\text{erase}} = 1 - (1-p_{\text{fail}}/2)\eta^{2}$.\\
The second approach to deal with the bias that we consider is to distribute it in alternating layers over the fusion network, which we call \textit{passive bias}.
In the low loss regime, this gives an increased tolerance to erasure errors, as the erasure errors from fusion failure can be seen to be restricted to spreading within
two-dimensional layers, for which the percolation threshold is higher than in three dimensions.
Given a bias in the erasure rates
\begin{equation}
\mathbf{B} = \frac{\text{min}(p_{\text{erase}}(ZZ),p_{\text{erase}}(XX))}{\text{max}(p_{\text{erase}}(ZZ),p_{\text{erase}}(XX))},
\end{equation}
a threshold value $\Tilde{p}_{\text{erase}}^{\mathbf{B}}$ for the worst-performing erasure rate is set.
Note that the worst-performing erasure rate is swapped in alternating layers between $XX$ and $ZZ$.
Both of the above biasing methods also apply to logical fusions,
where the $p_{\text{erase}}(XX)$ and $p_{\text{erase}}(ZZ)$ erasure rates are simply replaced by their logical counterparts $p_{\text{erase}}(\overline{XX})$ and $p_{\text{erase}}(\overline{ZZ})$.
The swapping of the biased logical erasure rate is described in Appendix.~\ref{App:dual_codes}.
\section{Logical fusion between identical graph codes details}
\label{app:losstol-logical-fusion}
We here introduce the details of the loss-tolerance and error correction of the logical fusions, starting with describing graph states and graph codes.
Graph states belong to the larger class of quantum states called stabilizer states \cite{gottesman1998}, with stabilizers generators given by $S_i = X_i \prod_{j \in \mathcal{N}(i)} Z_i$ where $i$ runs over all qubits in the graph state and $\mathcal{N}(i)$ indicates the neighborhood of qubit $i$ in the graph.
Products of these stabilizer generators generate the whole stabilizer group $\mathcal{S}$.
Turning a $n$ qubit graph state, denoted \textit{progenitor graph state} \cite{Tom}, into a $n-1$ qubit graph code initialized in the plus eigenstate of logical $X$, called $\ket{\overline{+}} $, is realized by choosing an input qubit $q$ and measuring it in the Pauli-$X$ basis obtaining a $+1$ outcome.
The logical operators of the graph code are then $\overline{X} = \prod_{i \in \mathcal{N}(q)} Z_i$ and $\overline{Z} = S_{q_0}Z_{q}$ for any choice of $q_0 \in \mathcal{N}(q)$ \footnote{The logical Y operator is given $\overline{Y} = i \overline{X}\overline{Z}$}, and its stabilizer group is retrieved from the stabilizers generators of the $n$-qubit graph as $\mathcal{S} = \langle S_{q_0}S_{i}, S_{j}  \rangle_{i \in \mathcal{N}(q)\setminus q_0, j \notin \mathcal{N}(q)}$.
Since the product of a logical operator and a stabilizer is still a valid logical operator, the full set of available logical operators are $\mathcal{L}^{X} = \overline{X} \cdot \mathcal{S}$ and $\mathcal{L}^{X} = \overline{Z} \cdot \mathcal{S}$.

\subsection{Erasure decoder}
\label{app:erasuredecoder}
As discussed in the main text, we consider logical fusion between two identical graph codes where equivalent code qubits are fused transversely.
For the physical fusions between code qubits, we consider the standard error model for non-boosted fusion \cite{PsiFBQC}:

\begin{equation*}
    \begin{tabular}{ |l||c|c|c|c|}
      \hline
      \makecell{Fusion\\ outcome} & \makecell{Success} & \makecell{Failure\\in Z} & \makecell{Failure\\in X} & \makecell{Photon\\loss} \\
      \hline
      \makecell{Measured\\parities} &  \makecell{$X_{1}X_{2}$\\$Z_{1}Z_{2}$}  & $X_{1}X_{2}$ & $Z_{1}Z_{2}$ & None  \\
      \hline
    \makecell{Probability}  & {($1-p_{\text{fail}})\eta^2$} & {$w p_{\text{fail}}\eta^2$} & {$(1-w)p_{\text{fail}}\eta^2$} & {$1 - \eta^2$} \\ 
      \hline
    \end{tabular}
\end{equation*}

Here, $1$ and $2$ label the two fusion qubits, $p_{\text{fail}} = \frac{1}{2}$ is the fusion failure probability for a non-boosted fusion \footnote{Strictly speaking, there are specific cases where a standard fusion can also succeed or fail with unit probability, e.g. if the two fused qubits are initially correlated \cite{Psiadaptive,fusionrules}}, and $\eta$ is the transmission efficiency seen by each photon. 
Failure in $X$ or in $Z$ is chosen by applying a Hadamard or not on the qubits before the fusion gate, which is indicated by the boolean parameter $w \in [0, 1]$. \\
For two graph codes with sets of logical operators $\mathcal{L}_{A/B}^{X}$ and $\mathcal{L}_{A/B}^{Z}$, a logical fusion succeeds if one pair of $\overline{X}_{A}\overline{X}_{B}$ and $\overline{Z}_{A}\overline{Z}_{B}$ is recovered after performing the physical fusions between all code qubits. 
Recovering only the $\overline{X}_{A}\overline{X}_{B}$ ($\overline{Z}_{A}\overline{Z}_{B}$) parity corresponds to a logical erasure of $\overline{Z}_{A}\overline{Z}_{B}$ ($\overline{X}_{A}\overline{X}_{B}$).
To calculate the logical erasure rates, we first group identical logical operators of the two codes $\mathcal{L} = \{\overline{X}^k_A\overline{X}^k_B \in \mathcal{L}_X^{A} \cdot \mathcal{L}_X^{B}, \quad \overline{Z}^k_A\overline{Z}^k_B \in \mathcal{L}_Z^{A} \cdot \mathcal{L}_Z^{B}\}$, where $k$ refers to the operator index.
From the fusion error model, we calculate all possible combinations of the physical fusion outcomes of all code qubits that recover a $\overline{X}^k_A\overline{X}^k_B$ and a $\overline{Z}^k_A\overline{Z}^k_B$~\cite{Tom}.
This represents a set of measurement patterns, which we shall denote
$\mathcal{M}_X$ and $\mathcal{M}_Z$.
Summing the probability of obtaining each measurement pattern in $\mathcal{M}_X$ and $\mathcal{M}_Z$ gives the success probability of recovering the parities
\begin{align}
    p_{\text{success}}(\overline{XX}) = \sum_{i \in \mathcal{M}_X} p_i, \\
    p_{\text{success}}(\overline{ZZ}) = \sum_{i \in \mathcal{M}_Z} p_i,
\end{align} 
where $p_i$ is the probability of obtaining measurement pattern $i$.
From the success probability the erasure rate is found from $p_{\text{erase}} = 1 - p_{\text{success}}$.
These logical erasure rates are a function of the photon loss $\gamma = 1 - \eta$ and the physical fusion failure bases set by $\mathbf{w} =\{w_i\}$, with $i$ running over all pairs of code qubits.
To find the best failure bases $\mathbf{w}$, we scan all $2^n$ possible combinations for two $n$-qubit graph codes and pick the failure basis configuration that allows for the largest $\gamma$ while the logical erasure rate is below the erasure threshold of the code given the implemented bias (i.e. $\Tilde{p}_{\text{erase}}$ for randomized bias, and $\Tilde{p}_{\text{erase}}^{\mathbf{B}}$ for passive bias).

\subsection{Pauli error decoder}
\label{app:error_decoder}

From the erasure decoder, described in Appendix.~\ref{app:erasuredecoder}, a set of measurement patterns $\mathcal{M}_X$ and $\mathcal{M}_Z$ which recover $\overline{X}_{A}\overline{X}_{B}$ and $\overline{Z}_{A}\overline{Z}_{B}$ are given.
In addition to retrieving the logical operators, each measurement pattern also gives access to a subset of the stabilizers of the two codes.
As for the logical operators, we only consider stabilizers that are formed from grouping identical stabilizers of the two codes $\mathcal{S} = \{\overline{S}^k_A\overline{S}^k_B \in \mathcal{S}^{A} \cdot \mathcal{S}^{B} \}$, where $\mathcal{S}^{A/B}$ is the stabilizer group of code $A$ and $B$ respectively. 
For a given successful trajectory, the subset $\mathcal{S_M} \in \mathcal{S}$ of available stabilizers are those that qubit-wise commute with the measurement pattern \cite{Tom}. With $\mathcal{S_M}$ we can perform error correction of the logical operators. \\
As an error model, we consider a single qubit depolarizing channel
\begin{equation}
    \mathcal{E} (\epsilon) = (1-\epsilon)\rho + \frac{\epsilon}{3}\sum_{\sigma \in \{ X, Y, Z\}}\sigma \rho \sigma,
\end{equation}
where $\epsilon$ is the single qubit error rate and $\rho$ is a single qubit density matrix.
As the qubits experience random Pauli errors the logical parities can flip, resulting in a logical error.
To calculate the probability of a logical error, we consider all possible combinations of Pauli errors on all pairs of fusion qubits.
For a depolarizing channel, the probability that the different parities (i.e. $XX$, $YY$ or $ZZ$) from the physical fusions are flipped is
\begin{equation}
    p = 4(\frac{\epsilon}{3}(1-\epsilon) + \frac{\epsilon^2}{9}),
\end{equation}
where $\epsilon/3$ is the probability of an $X$, $Y$ or $Z$ error on each fusion qubit.
From the sequence of Pauli errors, we check if the parities are flipped and extract the syndromes corresponding to the measured stabilizers.
For each syndrome, we determine the most likely error and correspondingly correct the logical parity outcome.
By summing the logical error rates for all possible syndromes weighted by their respective probability, we retrieve the logical error rate for the measured parity for a given measurement pattern.
This is done for all valid measurement patterns in $\mathcal{M}_X$ and $\mathcal{M}_Z$ for both logical parities, $\overline{X}_{A}\overline{X}_{B}$ and $\overline{Z}_{A}\overline{Z}_{B}$.
The total logical error rate for each parity is then the sum of the logical error rate for each measurement pattern weighted by its probability of occurring
\begin{equation}
    \label{eq:logical_error_rate}
    \overline{p}_{\text{error}}(\overline{\sigma}\overline{\sigma}) = \frac{\sum_{i \in \mathcal{M}_{\sigma}} p_i p_{i, \text{error}}(\overline{\sigma}\overline{\sigma})}{\sum_{i \in \mathcal{M}_{\sigma}} p_i}.
\end{equation}
Here, $\overline{\sigma}\overline{\sigma}$ indicates the measured logical parity, $p_i$ is the probability of obtaining the measurement pattern, and $p^i_{\text{error}}(\overline{\sigma}\overline{\sigma})$ is the logical error rate for the given measurement pattern. 
In the randomized basis architecture, given a logical erasure rate $\overline{p}_{\text{erase}}$ a fusion measurement error rate $\epsilon_M$ is inferred from the topological code which the average of the two logical error rates need to be below 
\begin{equation}
\label{eq:random_error_threshold}
\frac{\overline{p}_{\text{error}}(\overline{Z}\overline{Z}) + \overline{p}_{\text{error}}(\overline{X}\overline{X})}{2} < \epsilon_M.
\end{equation}
From Eq.~\ref{eq:random_error_threshold} a Pauli error threshold rate $\epsilon$ can be found, which is shown in Fig.~\ref{fig:results_error_loss} as a function of photon loss for the optimal codes.

\section{Dual inner codes}
\label{App:dual_codes}
Generating the dual inner code of a given graph code which swaps the erasure rates $p_{\text{erase}}(\overline{ZZ}) \leftrightarrow p_{\text{erase}}(\overline{XX})$, can be achieved by applying a Hadamard gate on the input qubit (i.e. the spin) of the progenitor graph.
This works since the stabilizers of the progenitor graph state with an $X$ or $Z$ on the input qubit (i.e. the spin) correspond to the $\overline{X}$ and $\overline{Z}$ of the graph code, respectively \cite{Tom}.
Thus, the Hadamard effectively swaps the logical operators, leaving the code stabilizers invariant.
Although the Hadamard gate presents an intuitive and easy way of generating the dual inner code, it also takes us out of the graph state space.
An alternative way of rotating the codes is through graph state transformation by applying Local complementation (LC)\cite{hein} on the progenitor graph of the graph code.
The action of LC on a node $q$ is to apply
\begin{equation}
    U^{LC}(q) = \sqrt{-iX_q}\prod_{j \in N(q)}\sqrt{iZ_j}
\end{equation}
to the graph state, and in a graphical picture transforms the induced subgraph of node $q$ to its complement \cite{hein}.
To see how the action of LC transforms the logical operators of the code, we look at how the Pauli operators transform under conjugation with $U^{LC}$, shown in table.~\ref{tab:LC_transform}.
The progenitor graph state of the dual inner code is found from a sequence of LC which transforms $X \leftrightarrow Z$ on the input qubit.
From table.~\ref{tab:LC_transform}, we see that this is achieved by the action of $U^{LC}(s) \otimes U^{LC}(q* \in N(s)) \otimes U^{LC}(s)$, where $s$ denotes the input qubit index.
Note, that all stabilizers transform according to Table.~\ref{tab:LC_transform} and leave their qubit support (i.e. the qubits which the stabilizers act on with a Pauli operator different from identity) invariant. 
Since the qubit support determines the loss tolerant properties of the logical fusion, applying $U^{LC}(s) \otimes U^{LC}(q* \in N(s)) \otimes U^{LC}(s)$ swaps the logical operators but leave their loss tolerant properties invariant \cite{Tom}.
The caveat to this invariance, is in the erasure decoder implementation, as we fix the failure basis to only $X$ or $Z$.
The sequence of LC could enforce the optimal failure basis to include $Y$ failure modes, and thus the decoder may change the loss-tolerant properties of the logical fusion.
However, as illustrated in Appendix.~\ref{app:graph_lib}, we found a dual inner code for all optimized codes in the main text this way. \\
\begin{table}[h]
    \centering
    \begin{tabular}{| m{7em} ||  m{1cm} m{1cm}  m{1cm}  m{1cm} |}
    \hline
 & \multicolumn{4}{c|}{Pauli operators } \\
        Qubit & X & Y & Z & $\mathbb{1}$ \\
        \hline
        $q* = q$ & X & -Z & Y & $\mathbb{1}$\\
        $q^* \in N(q)$ & -Y & X & Z & $\mathbb{1}$\\
        \hline
    \end{tabular}
    \caption{Transformations of the Pauli operators under conjugation with $U^{LC}(q)$, i.e. LC applied to qubit $q$. $q^*$ refers to a qubit index in the graph state.}
    \label{tab:LC_transform}
\end{table}

\section{Graph library}
\label{app:graph_lib}
\begin{figure*}[t!]
\includegraphics[width=1.0\textwidth]{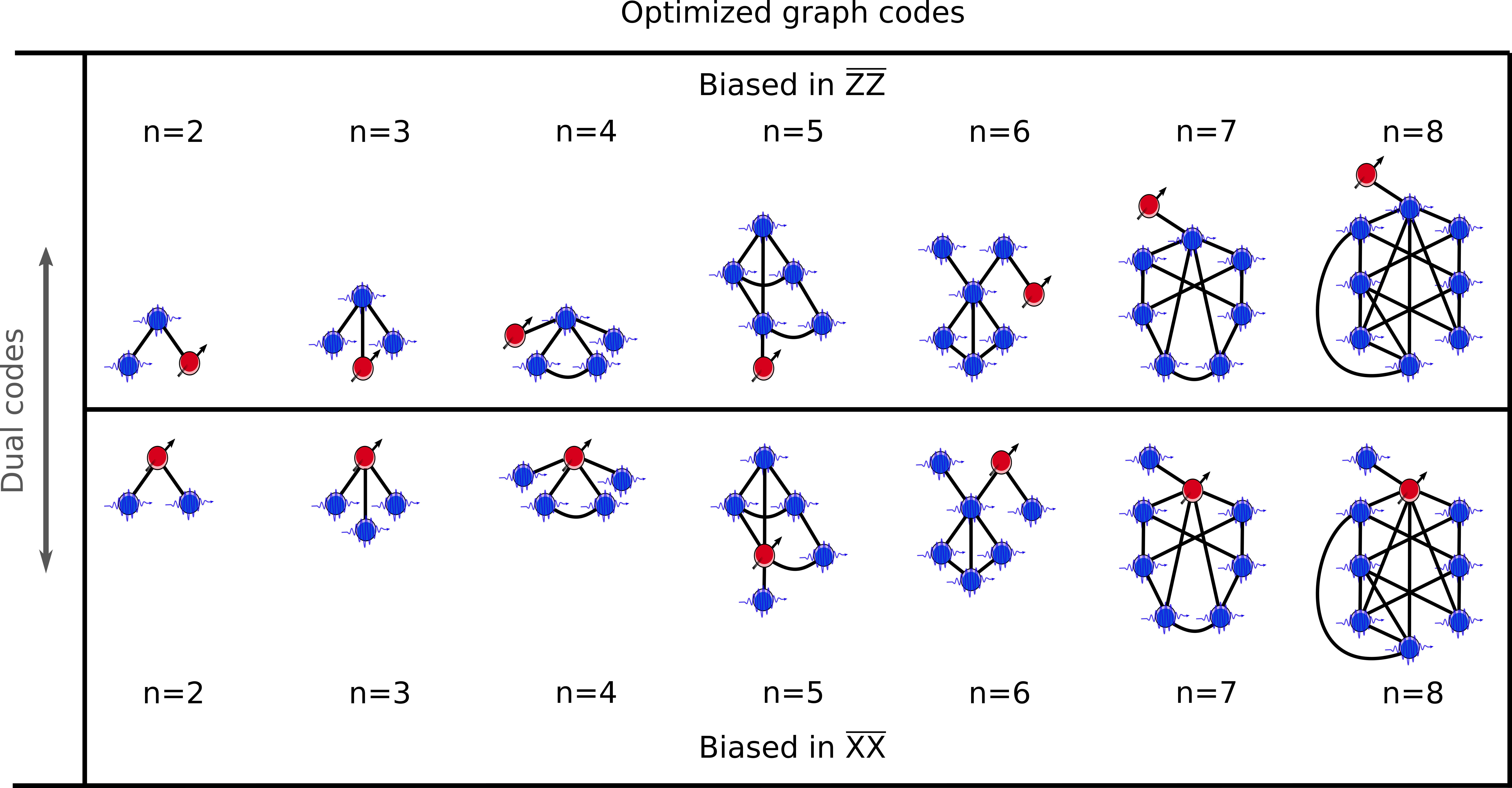}
\caption{\textbf{Graph code library}. The graph states to the optimized graph codes for loss-tolerance found in Fig.~\ref{fig:results}, where the graph state is turned into a graph code by measuring the spin (red) in X. \textit{Biased in $\overline{ZZ}$} are graph codes for which $p_{\text{erase}}(\overline{ZZ}) > p_{\text{erase}}(\overline{XX})$, and the \textit{Biased in $\overline{XX}$} are its dual counterpart where the erasure rate in $\overline{ZZ}$ and $\overline{XX}$ are swapped.
 \label{fig:graphlib}}
\end{figure*}

In Fig.~\ref{fig:graphlib} we illustrate the progenitor graph states and their dual partner corresponding to the best-performing graph codes for loss-tolerance.
The code pairs, primary and dual, are transformed between each other by the sequence of LC described in Appendix.~\ref{App:dual_codes}, which can be confirmed from Fig.~\ref{fig:graphlib}.
\end{document}